\documentclass[11pt]{article}
\usepackage{moriond,epsfig}

\def\be{\begin{equation}}
\def\ee{\end{equation}}
\def\bea{\begin{eqnarray}}
\def\eea{\end{eqnarray}}

\def\GeV{{\it A}GeV}
\def\fig#1{Fig.\ \ref{#1}}
\def\mt{\ensuremath{m_T}}
\begin{document}
\vspace*{4cm}
\title{ENERGY DEPENDENCE OF PARTICLE PRODUCTION IN NUCLEUS-NUCLEUS
  COLLISIONS AT THE CERN SPS}

\author{M. VAN LEEUWEN {\it for the NA49 collaboration}
  \footnote{presented at
    {\sc XXXVIII$^{th}$} Rencontres de Moriond on QCD and Hadronic
    interactions at high energy.}\\
  {\it NIKHEF, PO box 41882, 1009 DB Amsterdam,
  Netherlands }
  }
%\address{NIKHEF, PO Box 41882, 1009 DB Amsterdam, Netherlands}
\vfill{}
\maketitle
%\abstracts{
%No abstract yet.}

\section{Introduction}
From lattice QCD calculations, it is expected that hadronic matter
undergoes a phase transition to a deconfined state, the so-called
Quark Gluon Plasma (QGP), at an energy density of approximately
1~GeV$\;$fm$^{-3}$ or a temperature of 170~MeV.\cite{Karsch:2001vs}
The NA49 experiment is studying central lead-lead collisions at
different energies to search for signs of this transition. One of the
signals which is examined in this programme is the energy dependence
of strange particle production. New preliminary results on kaon and
pion production in central Pb+Pb collisions at 30~\GeV{} beam energy
are presented. These results are compared to the published NA49
data \cite{Afanasiev:2002mx} at 40, 80 and 158~\GeV{}
and to results from other experiments at lower and higher energies.
The data are then compared to expectations from models with and
without a phase transition to the QGP.

\section{Experiment and results}
The NA49 detector~\cite{Afanasev:1999iu} consists of four large Time
Projection Chambers (TPCs) which provide charged particle tracking and
particle identification through a measurement of the ionisation energy
loss $dE/dx$. Two of the TPCs are operated inside a magnetic field. In
addition, there are two Time-of-Flight (TOF) detectors, which improve
the particle identification capabilities at mid-rapidity.

\begin{figure}
  \begin{minipage}{0.65\textwidth}
  \epsfig{file=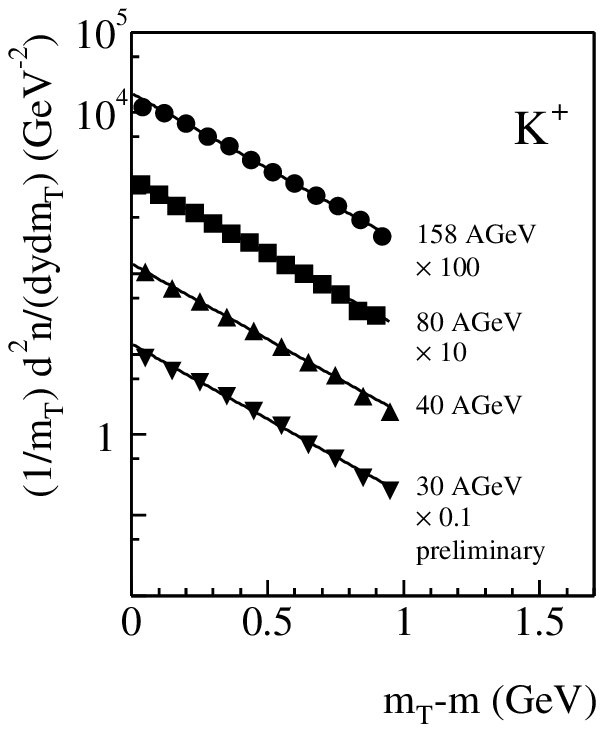,width=0.49\textwidth}
  %\end{minipage}
  %\hfill%
  %\begin{minipage}{0.4\textwidth}
  \epsfig{file=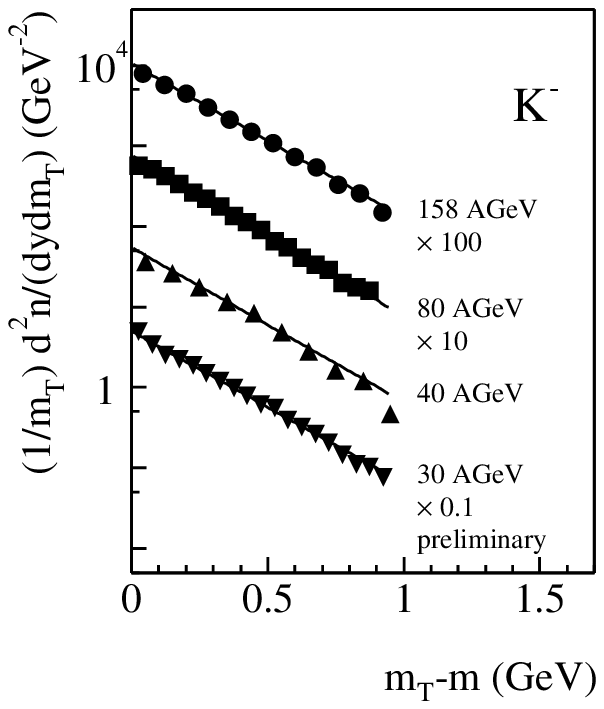,width=0.49\textwidth}
  %\vspace{-1cm}
  \caption{\label{fig:mtspec}Transverse mass spectra of $K^+$ (left)
    and $K^-$ (right) in central Pb+Pb collisions at four
    different energies. The lines are exponential fits to the data.}
  \end{minipage}
  \hfill%
  \begin{minipage}{0.33\textwidth}
    \epsfig{file=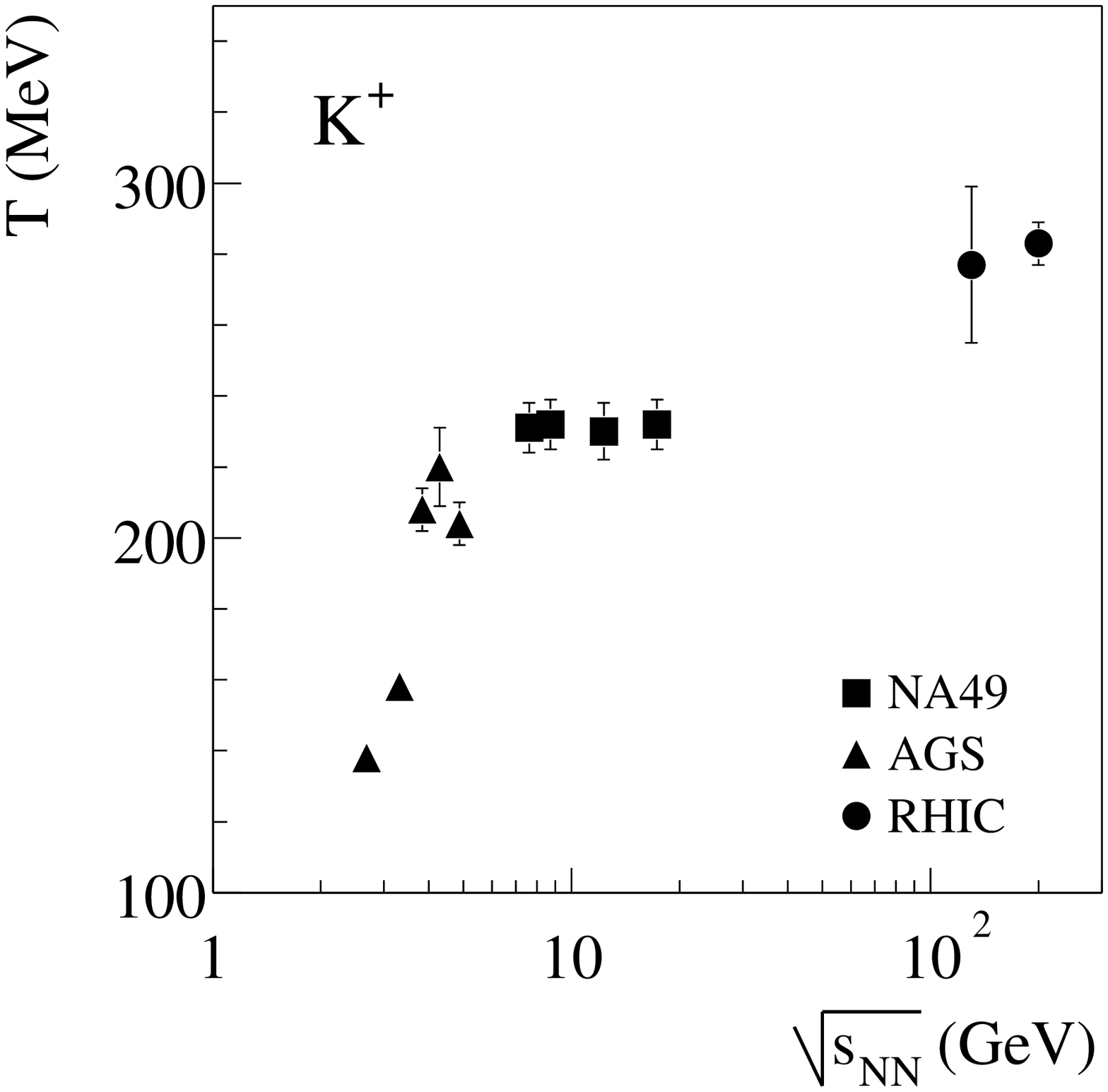,width=\textwidth}
    \caption{\label{fig:edep_mt}Energy dependence of the inverse slope
      parameters of $K^+$ \mt-spectra.}
  \end{minipage}  
\end{figure}
In \fig{fig:mtspec} we present, for the four available beam
energies, the transverse mass
($m_T=\sqrt{m^2+p_T^2}$) spectra of charged kaons at mid-rapidity, as
obtained from an analysis using the combined TOF and $dE/dx$
information for particle identification.

The lines indicate fits of an exponential distribution $dN/(\mt dy d\mt)
\propto \exp(-\mt/T)$. The energy dependence of the inverse slope
parameter $T$ of the $K^+$ spectra is shown in \fig{fig:edep_mt}. It
can be seen that this parameter strongly increases at AGS energies and
then stays constant in the SPS energy range from 30 to 158~\GeV.
\newpage 
\noindent There is some indication of a further increase towards RHIC energies.
The same behaviour is found for the inverse slopes of the $K^-$
spectra (not shown).\cite{Gorenstein:2003cu} 
In a hydrodynamic picture
of the collision, the increasing slope parameter is due to an
increase of the initial pressure with collision energy. The observed
energy dependence is qualitatively in agreement with a softening of
the equation of state due to a phase
transition.\cite{Gorenstein:2003cu,VanHove:1982vk}

\begin{figure}[hbt]
  \epsfig{file=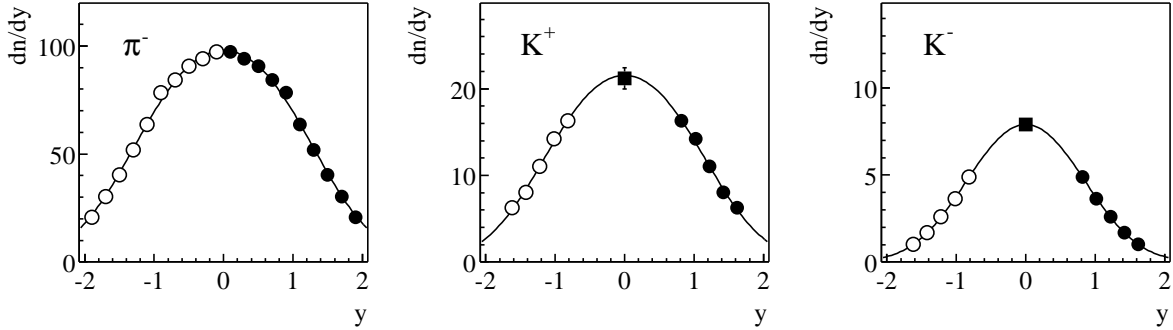,width=\textwidth}
  \vspace{-0.7cm}
  \caption{\label{fig:rap_spectra}Rapidity spectra of $\pi^-$ (left),
    $K^+$ (middle) and $K^-$ (right) in central Pb+Pb collisions at
    30~\GeV. The full symbols indicate measured values, while the open
    symbols are reflected around mid-rapidity. The lines are fits with
    a double Gaussian.}
    %The $\pi^-$
    %spectra are determined from all negatively charged particles,
    %subtracting feeddown from weak decays and the contribution of the
    %(measured) $K^-$ (see text).}
\end{figure}

The rapidity spectra of $\pi^-$, $K^+$ and $K^-$ at 30~\GeV{} are shown
in \fig{fig:rap_spectra}. The kaon yields were determined using the
combined TOF-$dE/dx$ information at mid-rapidity (squares) and $dE/dx$
information only at forward rapidities (circles). Due to their low
momenta, it is not possible to identify pions by $dE/dx$. The pion
yields were therefore determined from the spectra of negative
hadrons, subtracting contributions from $K^-$ and feeddown from weak
decays.  The rapidity spectra were fitted with a double Gaussian (full
line) to determine the total yields.

\section{Energy dependence of total yields}
\begin{figure}
\centering
\epsfig{file=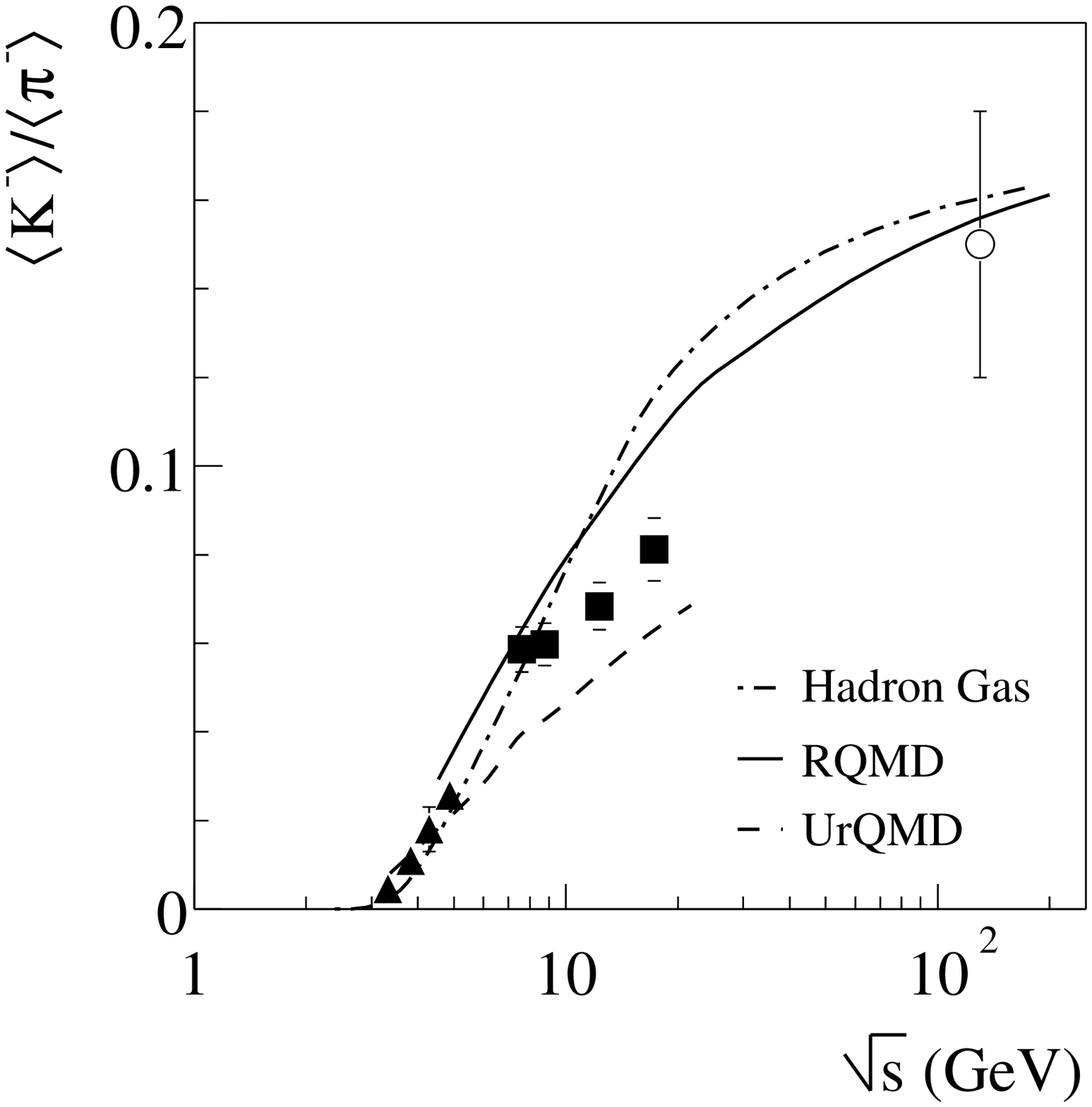, width=0.4\textwidth}%
\hspace{0.4cm}%
\epsfig{file=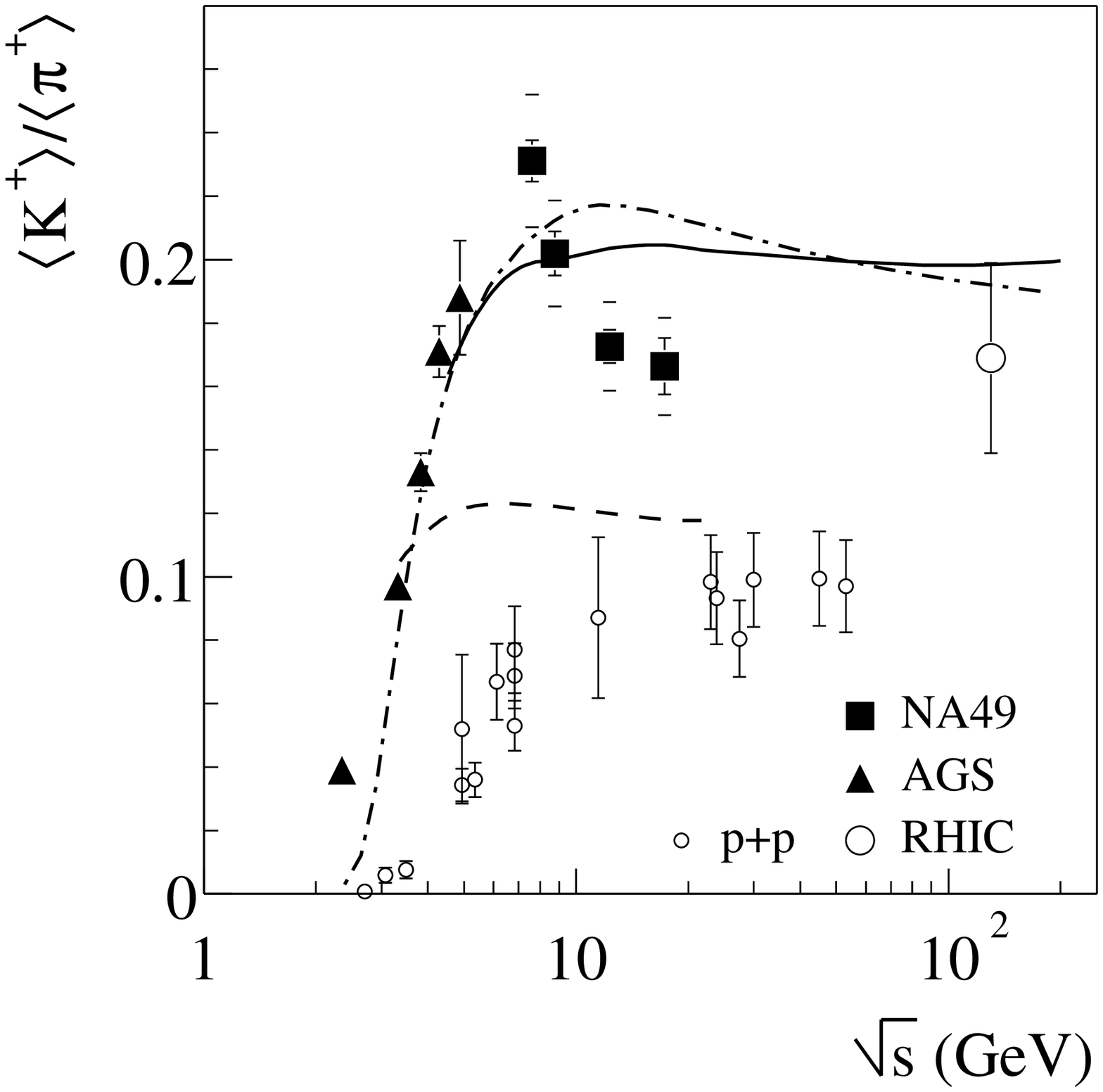, width=0.4\textwidth}
\vspace{-0.4cm}
\caption{\label{fig:edep_kapi}Energy dependence of the ratio of kaon and pion
  yields for positive and negative particles in nucleus-nucleus
  collisions and the
  $K^+/\pi^+$ ratio in proton-proton collisions. The lines show the
  expected energy dependence in three different models.}
\end{figure}
The energy dependence of the kaon to pion ratios are shown in
\fig{fig:edep_kapi}, where it can be seen that $K^-/\pi^-$ increases
monotonically with $\sqrt{s}$, while $K^+/\pi^+$ shows a relatively
sharp maximum at 30~\GeV{} and is constant at the higher SPS energies
(80 and 158~\GeV{}) up to RHIC energies. For comparison, the
$K^+/\pi^+$ ratio in proton-proton collisions is also shown. This
ratio is lower in proton-proton collisions than in nucleus-nucleus
collisions at all energies. This is true even at low energies, where
certainly no QGP formation is expected.  There is no indication
of a maximum in $K^+/\pi^+$ at 30~\GeV{} in proton-proton collisions,
contrary to nucleus-nucleus collisions.

The measured kaon to pion ratios are compared to different model
expectations as indicated by the lines in \fig{fig:edep_kapi}. Two of
these models, RQMD~\cite{Sorge:1995dp} (full line) and
UrQMD~\cite{Bleicher:1999xi,Weber:2002pk} (dashed line) are event
generators based on string excitation and decay. Secondary collisions
between produced particles are also taken into account. The relatively
large difference between the expectations of RQMD and UrQMD indicates
that there is a considerable uncertainty in such models.

The dash-dotted line in \fig{fig:edep_kapi} is the expectation from
the Hadron Gas Model. In this model, the
$K/\pi$ ratios are determined by only two thermodynamical
parameters: the temperature $T$ and the baryon-chemical potential
$\mu_B$.
The energy dependence is provided by a smooth parameterisation of the
energy dependence of $T$ and $\mu_B$, based on measurements at AGS and
the highest SPS energy. \cite{Cleymans:1999st}

A common feature of all three models is a smooth evolution of $K/\pi$
with energy. 
Only in the Hadron Gas Model a maximum in the
$K^+/\pi^+$ ratio is expected at low SPS energies. This
maximum, however, is much
less pronounced than 
%The energy dependence of the $K^+/\pi^+$ ratio in the Hadron Gas
%Model has a maximum at low SPS energies in the model, but the maximum
%is much less pronounced than 
observed in the data.

\begin{figure}[bt]
\begin{minipage}[t]{0.48\textwidth}
\centering
\epsfig{file=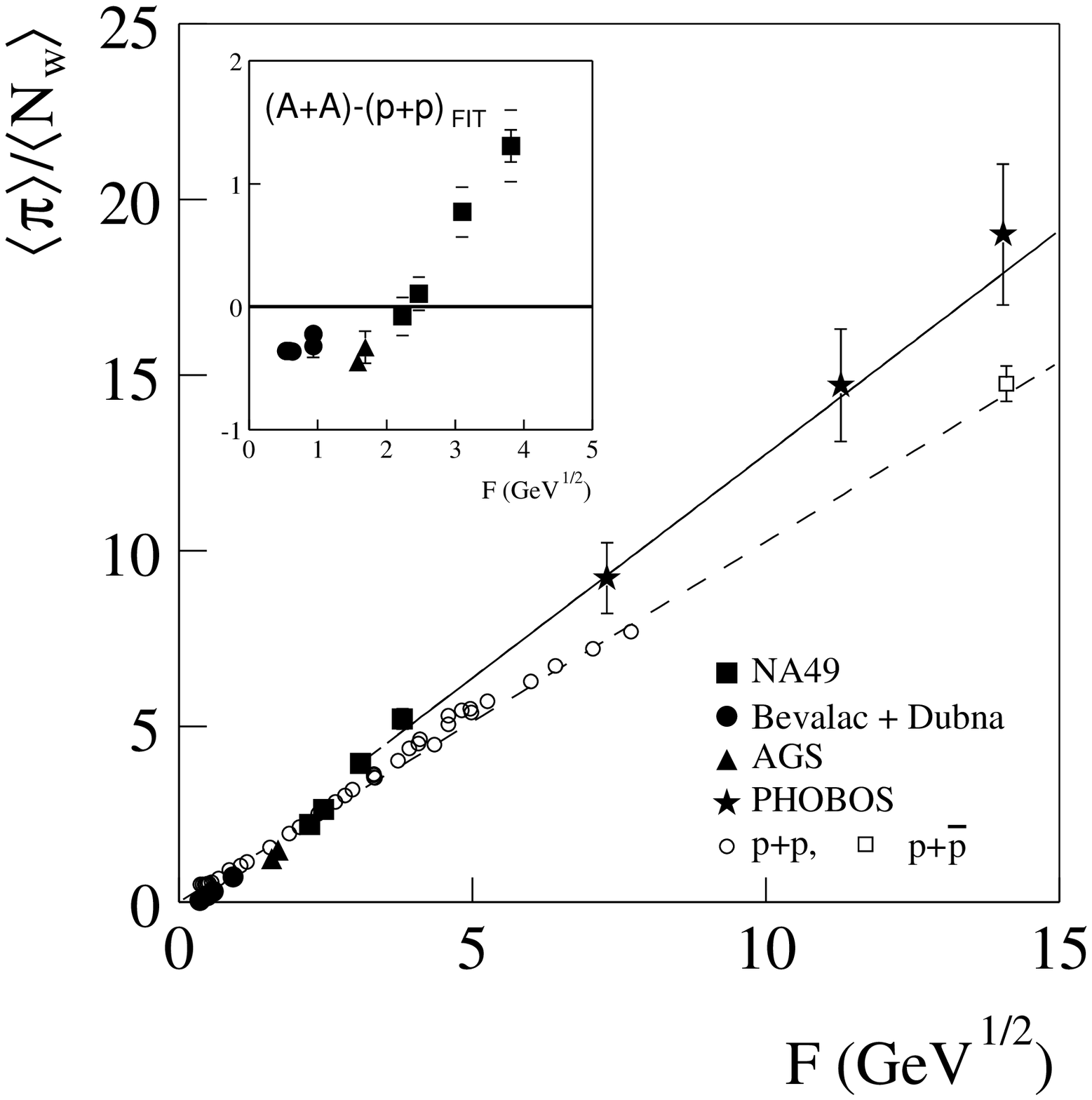, width=0.8\textwidth}
\vspace{-0.2cm}
\caption{\label{fig:pinp}Dependence of the pion multiplicity per
  wounded nucleon on the collision energy $F\approx s^{1/4}$ (see
  text). The inset shows the difference between
  the measurements in nucleus-nucleus collisions and the parametrised
  energy dependence in proton-proton collisions (open circles, dashed line).}
\end{minipage}
\hfill%
\begin{minipage}[t]{0.48\textwidth}
\centering
\epsfig{file=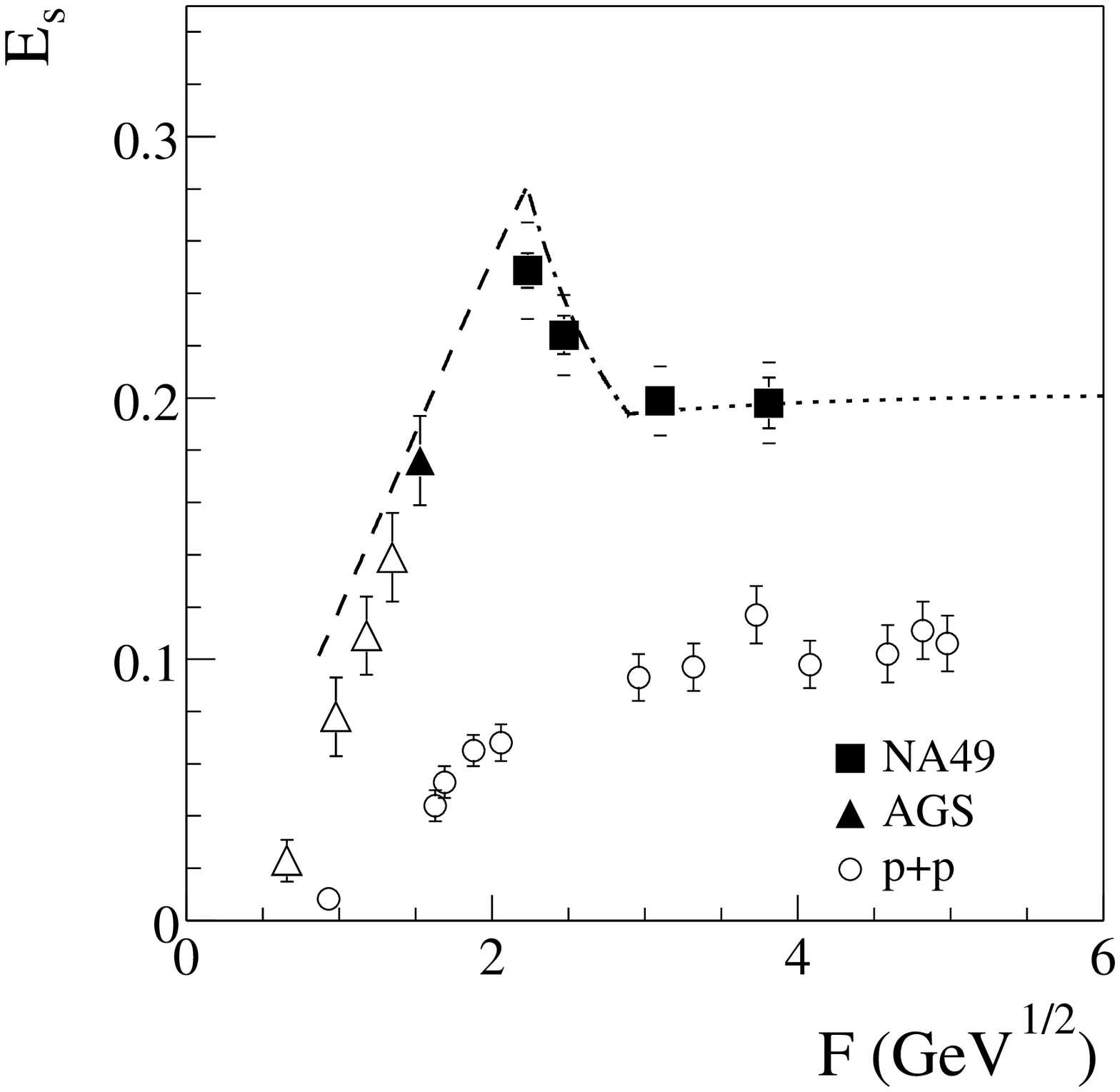, width=0.8\textwidth}
\vspace{-0.2cm}
\caption{\label{fig:es}Dependence of the strangeness to pion ratio
  $E_s$ on the collision energy $F\approx s^{1/4}$ (see text). Also
  shown is the expectation within the Statistical Model of The Early
  Stage (lines).\protect\cite{Gazdzicki:1998vd}}
\end{minipage}
%\vspace{-0.2cm}
\end{figure}

In Figs.~\ref{fig:pinp} and \ref{fig:es}, the experimental data are
compared to the Statistical Model of the Early Stage
(SMES) \cite{Gazdzicki:1998vd} in which it is assumed that a phase transition
to the QGP occurs.

In \fig{fig:pinp} the number of produced pions $\langle\pi\rangle=1.5
(\langle \pi^- \rangle + \langle \pi^+ \rangle)$ per wounded nucleon
is shown as a function of collision energy, characterised by
$F\equiv(\sqrt{s}-2m_N)^{3/4}/\sqrt{s}^{1/4}$, where $m_N$ is the
nucleon mass. Within the SMES, the pion multiplicity per wounded
nucleon is proportional to the initial entropy per wounded nucleon,
which, in turn, is proportional to $F$. The change of slope of the
energy dependence of the $\langle \pi \rangle/N_W$ ratio in
nucleus-nucleus collisions (inset of \fig{fig:pinp}) would then
indicate an increase of the initial number of degrees of freedom in
the collision.

In \fig{fig:es}, the energy dependence of the total strangeness to
pion ratio $E_s=( \langle \Lambda \rangle + 2 (\langle K^+ \rangle +
\langle K^- \rangle ))/\langle\pi\rangle$ is compared to the
expected ratio in the SMES. The lambda yields were estimated as
$\langle\Lambda\rangle=(\langle K^+\rangle - \langle K^- \rangle)/0.8$
since no published data are available. Within the SMES, the sharp peak
in the strangeness to entropy (or pion) ratio is due to the phase
transition which is assumed to set in at a temperature of $T=200$~MeV
(or $F\approx2.2$~GeV$^{1/2}$). The strangeness to entropy ratio in
the pure QGP phase, above $F\approx2.7$~GeV$^{1/2}$, is a
parameter-free prediction because it is determined by the ratio of the
number of strange degrees of freedom to the total number of degrees of
freedom.

To summarise, we observe that the kaon slope parameters in
nucleus-nucleus collisions show an overall increase with collision
energy, but are energy-independent at the SPS. Furthermore, the
$K^+/\pi^+$ ratio shows a pronounced maximum at about 30~\GeV{} which
seems to be specific to nucleus-nucleus collisions.  Qualitatively,
these observations can be understood by assuming that a phase
transition occurs between the highest AGS energy and 30~\GeV{} (SMES
model), while, at present, they cannot be accommodated in hadronic
descriptions of the collision such as RQMD, UrQMD and the Hadron Gas
Model.

\section*{Acknowledgments}
{\small
This work was supported by the Director, Office of Energy Research, Division of Nuclear Physics of the Office of
High Energy and Nuclear Physics of the US Department of Energy (DE-ACO3-76SFOOO98 and DE-FG02-91ER40609), the US National
Science Foundation, the Bundesministerium fur Bildung und Forschung, Germany, the Alexander von Humboldt Foundation, the UK
Engineering and Physical Sciences Research Council, the Polish State Committee for Scientific Research (2 P03B 130 23,
SPB/CERN/P-03/Dz 446/2002-2004, 2 P03B 02418, 2 P03B 04123), the Hungarian Scientific Research Foundation (T032648, T14920 and
T32293), Hungarian National Science Foundation, OTKA, (F034707), the
EC Marie Curie Foundation, and the Polish-German Foundation. }


\section*{References}
\small
\begin{thebibliography}{99}
%\cite{Karsch:2001vs}
\bibitem{Karsch:2001vs}
F.~Karsch,
%``Lattice results on QCD thermodynamics,''
Nucl.\ Phys.\ A {\bf 698} (2002) 199
[arXiv:hep-ph/0103314].
%%CITATION = HEP-PH 0103314;%%

%\cite{Afanasiev:2002mx}
\bibitem{Afanasiev:2002mx}
S.~V.~Afanasiev {\it et al.}  [The NA49 Collaboration],
%``Energy dependence of pion and kaon production in central Pb + Pb  collisions,''
Phys.\ Rev.\ C {\bf 66} 054902 (2002) 
[arXiv:nucl-ex/0205002].
%%CITATION = NUCL-EX 0205002;%%

%\cite{Afanasev:1999iu}
\bibitem{Afanasev:1999iu}
S.~Afanasev {\it et al.}  [NA49 Collaboration],
%``The NA49 large acceptance hadron detector,''
Nucl.\ Instrum.\ Meth.\ A {\bf 430}, 210 (1999).
%%CITATION = NUIMA,A430,210;%%

%\cite{Gorenstein:2003cu}
\bibitem{Gorenstein:2003cu}
M.~I.~Gorenstein, M.~Gazdzicki and K.~A.~Bugaev,
%``Transverse activity of kaons and the deconfinement phase transition in  nucleus nucleus collisions,''
arXiv:hep-ph/0303041.
%%CITATION = HEP-PH 0303041;%%

%\cite{VanHove:1982vk}
\bibitem{VanHove:1982vk}
L.~Van Hove,
%``Multiplicity Dependence Of P(T) Spectrum As A Possible Signal For A  Phase Transition In Hadronic Collisions,''
Phys.\ Lett.\ B {\bf 118}, 138 (1982).
%%CITATION = PHLTA,B118,138;%%

%\cite{Sorge:1995dp}
\bibitem{Sorge:1995dp}
H.~Sorge,
%``Flavor Production in Pb(160AGeV) on Pb Collisions: Effect of Color Ropes and Hadronic Rescattering,''
Phys.\ Rev.\ C {\bf 52} (1995) 3291
[arXiv:nucl-th/9509007].
%%CITATION = NUCL-TH 9509007;%%

%\cite{Bleicher:1999xi}
\bibitem{Bleicher:1999xi}
M.~Bleicher {\it et al.},
%``Relativistic hadron hadron collisions in the ultra-relativistic quantum  molecular dynamics model,''
J.\ Phys.\ G {\bf 25} (1999) 1859
[arXiv:hep-ph/9909407].
%%CITATION = HEP-PH 9909407;%%

%\cite{Weber:2002pk}
\bibitem{Weber:2002pk}
H.~Weber, E.~L.~Bratkovskaya, W.~Cassing and H.~Stocker,
%``Hadronic observables from SIS to SPS energies: Anything strange with  strangeness?,''
Phys.\ Rev.\ C {\bf 67} (2003) 014904
[arXiv:nucl-th/0209079].
%%CITATION = NUCL-TH 0209079;%%

%\cite{Cleymans:1999st}
\bibitem{Cleymans:1999st}
J.~Cleymans and K.~Redlich,
%``Chemical and thermal freeze-out parameters from 1-A-GeV to 200-A-GeV,''
Phys.\ Rev.\ C {\bf 60}, 054908 (1999)
[arXiv:nucl-th/9903063].
%%CITATION = NUCL-TH 9903063;%%

%\cite{Gazdzicki:1998vd}
\bibitem{Gazdzicki:1998vd}
M.~Gazdzicki and M.~I.~Gorenstein,
%``On the early stage of nucleus nucleus collisions,''
Acta Phys.\ Polon.\ B {\bf 30}, 2705 (1999)
[arXiv:hep-ph/9803462].
%%CITATION = HEP-PH 9803462;%%

\end{thebibliography}
\end{document}